\documentclass[aps,prb,reprint,twocolumns,superscriptaddress,nofootinbib]{revtex4-2}
\usepackage[utf8]{inputenc}

\usepackage[hidelinks]{hyperref}

\usepackage{hyperref}
\usepackage{color}
\hypersetup{colorlinks=true}
\usepackage{graphicx}
\usepackage{bm}
\usepackage{amsmath}
\usepackage{amssymb}
\usepackage{chemformula}
\usepackage[capitalise]{cleveref}
\usepackage{siunitx}
\usepackage{txfonts}
\newcommand{\hc}{\text{h.c.}}
\newcommand{\ii}{\mathrm{i}}

\renewcommand{\Im}{\operatorname{Im}}

\DeclareMathOperator{\sign}{sign}

\newcommand{\subfigref}[2]{Fig.~\hyperref[#1]{\ref*{#1}#2}}

\newcommand{\Jexchangecupr}{\SI{1.0}{meV}}
\newcommand{\Kanicupr}{\SI{0.01}{meV}}

\newcommand{\epsilonplasmon}{\num{10}}
\newcommand{\gammaplasmon}{\num{e-02}}
\newcommand{\Efermi}{\SI{100}{meV}}
\newcommand{\Jexchange}{\SI{0.75}{meV}}
\newcommand{\Kani}{\SI{0.05}{meV}}
\newcommand{\gpolarizationrutile}{\num{e-02}}

\begin{document}
	\title{Coupling of plasmons to the two-magnon continuum in antiferromagnets}
	\date{\today}
	
	\author{Pieter M. Gunnink}
	\email{pgunnink@uni-mainz.de}
	\affiliation{Institute of Physics, Johannes Gutenberg-University Mainz, Staudingerweg 7, Mainz 55128, Germany}
	\author{Alexander Mook}
	\affiliation{Institute of Physics, Johannes Gutenberg-University Mainz, Staudingerweg 7, Mainz 55128, Germany}
	\begin{abstract}   
		The coupling of magnons and plasmons offers a promising avenue for hybrid quantum systems, facilitating coherent energy and information transfer between magnetic and charge excitations. However, existing mechanisms often depend on spin-orbit coupling or temperature-activated processes, limiting their robustness for low-temperature quantum technologies. Here, we propose a coupling mechanism between plasmons and the two-magnon continuum in antiferromagnetic insulators, which operates at zero temperature and does not require spin-orbit coupling. Using a model system consisting of a two-dimensional electron gas on an insulating antiferromagnetic substrate, we show that the electric field of the plasmons interacts with the magnetically mediated electric polarization in the antiferromagnet, arising from bonds with broken inversion symmetry. This interaction enables a strong coupling to the spin-conserving two-magnon continuum, allowing for efficient hybridization and reaching the ultrastrong coupling regime.
	\end{abstract}
	
	\maketitle

	\section{Introduction}
	
	Magnonics is an alternative approach to conventional charge-based electronic computing, bypassing the Joule heating associated with electronic devices \cite{flebus2024MagnonicsRoadmap2024}. Within the field of magnonics, antiferromagnetic materials stand out from a technological perspective \cite{baltzAntiferromagneticSpintronics2018}, due to their natural high frequencies in the terahertz range. To further develop antiferromagnets as a materials platform for next-generation computing, it is of great importance to develop interoperable magnonics devices---to interface with the wide range of quasiparticles currently being investigated as possible information carriers in condensed matter, photonics and other fields \cite{zarerameshtiCavityMagnonics2022,yuanQuantumMagnonicsWhen2022,chumakMagnonSpintronics2015}. 
	
	One of the possible information carriers considered are plasmons, the quanta of charge-carrier density oscillations \cite{alonsocalafellQuantumComputingGraphene2019}. 
	The emergence of two-dimensional (2D) materials that can support gapless plasmon modes, such as graphene \cite{hwangDielectricFunctionScreening2007, vafekThermoplasmaPolaritonScaling2006, wangCollectiveExcitationsDirac2007, wunschDynamicalPolarizationGraphene2006,luoPlasmonsGrapheneRecent2013}, has opened up the possibility of magnon-plasmon coupling, and previous works have already demonstrated the coupling between magnons and plasmons \cite{efimkinTopologicalSpinplasmaWaves2021, dyrdalMagnonplasmonHybridizationMediated2023,costaStronglyCoupledMagnonplasmon2023, ghoshPlasmonmagnonInteractionsTwodimensional2023,yuanStrongTunableCoupling2024,yuanBreakingSurfacePlasmonExcitation2024}. These works considered either metallic 2D magnets, hosting both magnons and plasmons \cite{ghoshPlasmonmagnonInteractionsTwodimensional2023,dyrdalMagnonplasmonHybridizationMediated2023}, or heterostructures of a 2D metal and an insulating magnet \cite{efimkinTopologicalSpinplasmaWaves2021,costaStronglyCoupledMagnonplasmon2023,yuanStrongTunableCoupling2024,yuanBreakingSurfacePlasmonExcitation2024}. The primary focus has been on bilinear magnon-plasmon coupling, hybridizing magnons and plasmons. This can be achieved through the Zeeman coupling of the electromagnetic field of the plasmon with the spins \cite{costaStronglyCoupledMagnonplasmon2023,yuanStrongTunableCoupling2024,yuanBreakingSurfacePlasmonExcitation2024}, or through spin-orbit coupling, which couples the plasmons directly to the spin degree of freedom \cite{dyrdalMagnonplasmonHybridizationMediated2023,efimkinTopologicalSpinplasmaWaves2021}. Beyond bilinear plasmon-magnon coupling, Ref.~\cite{ghoshPlasmonmagnonInteractionsTwodimensional2023} has shown that in a metallic 2D honeycomb-lattice ferromagnet, the plasmonic excitations can couple---even without spin-orbit coupling---to the thermally activated magnon density.

	
	This raises the question of whether such coupling can also occur in the absence of spin-orbit and magnetic dipolar interactions, and at zero temperature. If so, it could potentially exceed the intrinsic decay rate of the quasiparticles, offering a promising avenue for quantum technologies operating at ultralow temperatures. In this work, we provide an affirmative answer. We show that plasmons can also couple to the \emph{continuum} formed by antiferromagnetic (AFM) magnons. This coupling is enabled by the electric field of the plasmon that interacts with the microscopic electric polarization of the material and, in turn, with the magnetic order \cite{bolensTheoryElectronicMagnetoelectric2018}. 
	Importantly, this plasmon-magnon continuum coupling does not require spin-orbit coupling, since in an AFM there is a magnon continuum built from two magnons with opposite spin, such that spin is conserved.
	In addition, the coupling remains at zero temperature. We argue that it naturally holds the potential for realizing a large coupling rate between magnons and plasmons, sufficient to reach the ultrastrong coupling regime \cite{friskkockumUltrastrongCouplingLight2019}, where the coupling exceeds the typical magnon and plasmon decay rate. This development opens the way for coherent transfer of quantum information \cite{kurizkiQuantumTechnologiesHybrid2015,friskkockumUltrastrongCouplingLight2019} between magnons and plasmons \cite{jiangIntegratingMagnonsQuantum2023, liHybridMagnonicsPhysics2020,yuanQuantumMagnonicsWhen2022}. 
	
	\begin{figure}[b]
		\centering
		\includegraphics[width=\columnwidth]{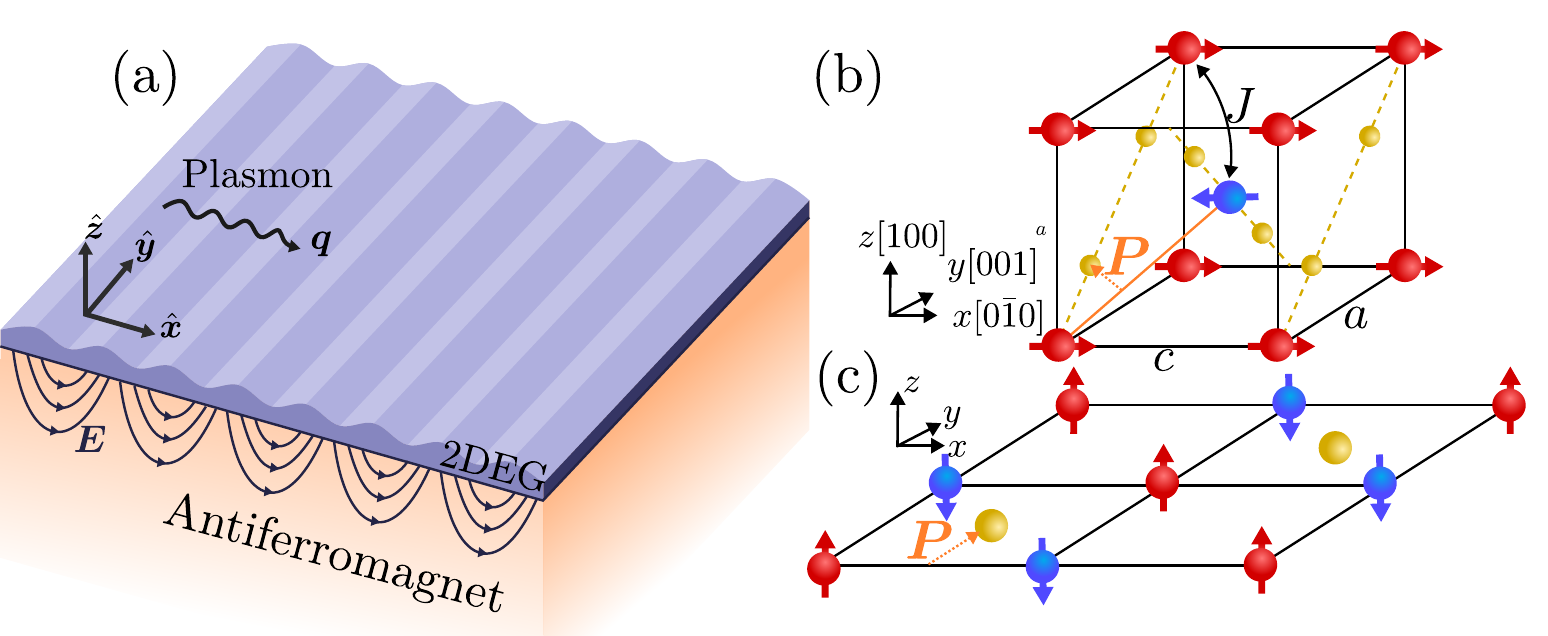}
		\caption{(a) The setup considered, where a two-dimensional electron gas (2DEG) is placed on top of a semi-infinite insulating antiferromagnet. The 2DEG supports gapless plasmons, and their electric field extends into the antiferromagnet, where they couple to the two-magnon continuum.  (b) A three-dimensional antiferromagnet inspired by the rutile-type antiferromagnets, such as \ch{MnF2}, grown along [100]. The magnetic ions (red and blue) are antiferromagnetically ordered, and the anions (yellow) break the inversion symmetry along the nearest-neighbor bonds. This allows for a finite polarization vector $\bm P$, as indicated for one bond. (c) A quasi two-dimensional antiferromagnet. The magnetic ions  are antiferromagnetically ordered and form a square lattice, and the non-magnetic ions (yellow) break the center of inversion symmetry located between neighbhoring magnetic ions.  \label{fig:setup}}
	\end{figure}
	
	To demonstrate this coupling, we consider plasmons in a two-dimensional electron gas (2DEG), placed on top of an antiferromagnetic crystal, as shown in \cref{fig:setup}. The plasmons generate an electric field \cite{ferreiraQuantizationGraphenePlasmons2020}, which extends into the insulating antiferromagnetic crystal and couples to the magnetic excitations. 
	Similar coupling schemes have been previously considered in cuprates placed in a cavity \cite{curtisCavityMagnonpolaritonsCuprate2022} or driven by a laser \cite{bossiniLaserdrivenQuantumMagnonics2019,bossiniUltrafastAmplificationNonlinear2021,shanDynamicMagneticPhase2024}. However, magnonic energies in cuprates are much higher (in the \SIrange{100}{500}{meV} range) than magnonic energies in other antiferromagnets (in the \SIrange{1}{10}{meV} range), but cavities and lasers operating at these lower energy scales are much harder to realize. In this work, we therefore propose to use plasmons to bridge this gap.

	In order for the magnetic excitations to couple linearly to electric fields, the microscopic polarization operator in the antiferormagnet should be non-zero \cite{sergienkoFerroelectricityMagneticEPhase2006,bolensTheoryElectronicMagnetoelectric2018}. In absence of spin-orbit coupling, it can be shown from symmetry considerations that this is only possible if the center of a bond connecting two atoms is not a center of inversion. We stress that this does not imply a broken global inversion symmetry, nor the existence of ferroelectricity in the ground state. We choose here two systems to illustrate the magnon-plasmon coupling, as shown in \cref{fig:setup}: (i) a rutile-type antiferromagnet, where the anions naturally break the bond inversion symmetry, and (ii) a layered quasi two-dimensional antiferromagnet, where we break the bond inversion symmetry by placing non-magnetic ions in a checkerboard-like fashion.

	This article is layout as follows. In \cref{sec:polarization} we first discuss the coupling between an electric field and spin excitations in an insulating magnetic material. Next, in \cref{sec:method} we present our main result: the coupling between magnons and plasmons, which gives rise to a plasmon self-energy. In \cref{sec:results} we show this plasmon self-energy and the resulting plasmon spectral function for the two systems we consider. We end with a conclusion and discussion in \cref{sec:conclusion}.

	\section{Polarization}
	\label{sec:polarization}
	
	In this work, we focus on the coupling of the magnetic excitations to an electric field \cite{damascelliDirectTwoMagnonOptical1998,azumaOpticalResponseDue2005,schoenfeldDynamicalRenormalizationMagnetic2024}. This coupling is described by the polarization operator, $\hat{\bm P_{i}}$, which results in the Hamiltonian \cite{moriyaTheoryLightScattering1967,zhuSpinorbitCouplingElectronic2014,malzTopologicalMagnonAmplification2019,seifertOpticalExcitationMagnons2019,mostovoyMultiferroicsDifferentRoutes2024}
	\begin{equation}
		\mathcal H_c = -\sum_i \hat{\bm E}_i \cdot \hat{\bm P_{i}}
	\end{equation}
	where 
	\begin{equation}
		\hat{\bm P_{i}} = \sum_j\left(\bm\pi_{ij}\delta_{\alpha\beta} +\bm\Gamma_{ij}^{\alpha\beta} \right) \hat S_i^\alpha \hat S_j^\beta
	\end{equation}
	is the polarization operator expanded up to second order in spin operators. 
	We have decomposed the polarization tensor into the isotropic part $\bm \pi$ and the traceless part $\bm \Gamma$. The allowed components of this tensor are determined by lattice symmetries, and are modified in the presence of spin-orbit coupling. 
	In this work we consider only the isotropic part $\bm\pi$, which does not require spin-orbit coupling and is non-zero for any bond where its center is not a center of inversion. Microscopically, this is the exchange-striction mechanism arising from the symmetric spin exchange interaction \cite{sergienkoFerroelectricityMagneticEPhase2006,bolensTheoryElectronicMagnetoelectric2018}
	
	The direction of $\bm \pi$ is set by the crystal symmetries \cite{moriyaTheoryLightScattering1967,seifertOpticalExcitationMagnons2019,bolensTheoryElectronicMagnetoelectric2018}, whereas its magnitude depends on the microscopic details. Generally, the polarization $\bm\pi_{ij}$ between sites $i,j$ is mediated by an additional site $k$ which mediates the hopping and breaks the inversion symmetry for this specific bond. One can then write
	$  \bm \pi_{ij} = g_{ij}a_{ij}(\bm e_{jk} - \bm e_{ki})$ \cite{zhuSpinorbitCouplingElectronic2014}, where $g$ is a parameter describing the polarization amplitude, $a_{ij}$ is the bond length and $\bm e_{ij}$ is the unit vector between atoms $i$ and $j$. Within a Fermi-Hubbard model at half filling it is possible to derive the polarization amplitude as 
	$g_{ij}= 8e t_{ij}t_{jk}t_{ki}/U^3$, where $e$ is the electron charge, $t_{ij}$ is the hopping amplitude between $i$ and $j$, and $U$ is the Coulomb repulsion \cite{zhuSpinorbitCouplingElectronic2014}. Note that this expression was derived in the Mott insulating limit, and therefore $t/U\ll1$. In this work, we will consider the polarization amplitude $g_{ij}$ to be a phenomenological parameter.

	We continue by considering the electric field operator $\hat{\bm E}_i$ of a plasmon in the 2DEG, which at position $\bm r_i$ in the antiferromagnet is given by \cite{ferreiraQuantizationGraphenePlasmons2020}
	\begin{equation}
		\hat{\bm E}(\bm r_i)= \frac{1}{\sqrt{N_p}}\sum_{\bm q} \ii \bm E_{\bm q}e^{i\bm q \cdot \bm r_i- |z_i|/\lambda_{\bm q}} \hat\phi_{\bm q} + \hc\, ,
	\end{equation}
	with $\hat\phi_{\bm q}$ the plasmon annihilation operator and $\bm E_{\bm q}(z)$ a mode function, which captures the details of the plasmonic dispersion. We give $\bm E_{\bm q}(z)$ and further details of the plasmons in the 2DEG in \cref{app:plasmon}. Here $N_p$ is the number of atomic sites in the 2DEG and the plasmon electric field decays exponentially along the $z$-axis, with a decay length $\lambda^{-1}\approx q$ for the parameters considered here.

	Continuing with only the isotropic part $\bm \pi$ of the polarization tensor, the coupling Hamiltonian can then be written as
	\begin{equation}
		\mathcal H_c = -\sum_{ij} (\hat{\bm E}(\bm r_i) \cdot \bm\pi_{ij}) \hat{\bm S}_i \cdot \hat{\bm S}_j, \label{eq:Hc-pi}
	\end{equation}
	where we have used $\hat{\bm E}(\bm r_j) \approx \hat{\bm E}(\bm r_i)$, since $\bm\pi_{ij}$ is only finite for two sites $i$ and $j$ so close to each other that the spatial modulation of the electric field is negligible.

	\section{Magnon-plasmon coupling}
	\label{sec:method}
	We consider here an antiferromagnet described by the Heisenberg Hamiltonian,
	\begin{equation}
		\mathcal H_{\mathrm{AFM}} = \sum_{ij}J_{ij}\hat{ \bm S}_i \cdot \hat{ \bm S}_j - \frac{K}{2}\sum_i (\hat S^z_i)^2, \label{eq:HAFM}
	\end{equation}
	where $J_{ij}$ is the Heisenberg exchange strength between spins at site $i,j$ and $K>0$ parameterizes an easy-axis anisotropy. We assume that the ground state of the system is the antiferromagnetic Néel state, such that its elementary magnetic excitations are magnons. To describe the magnonic excitations, we employ a Holstein-Primakoff transformation, drop the many-body terms, apply a Fourier transformation and
	a Bogoliubov transformation (see \cref{app:hamiltonian} for the details), such that the Hamiltonian in \cref{eq:HAFM} becomes 
	\begin{equation}
		H=\sum_{\bm k} \epsilon_{\bm k} \left(  \hat \alpha_{\bm k}^\dagger \hat \alpha_{\bm k} + \hat\beta_{\bm k}^\dagger \hat\beta_{\bm k} \right),
	\end{equation}
	where $\epsilon_{\bm k}$ is the magnon eigenenergy, given in \cref{app:hamiltonian}, and we have dropped the quantum fluctuations.
	
	We will consider in this work two examples of antiferromagnets with non-zero electric-dipole coupling: (i) a  three-dimensional antiferromagnet and (ii) a quasi-two dimensional layered antiferromagnet, both shown in \cref{fig:setup}. As a model for the 3D AFM we take an easy-axis rutile antiferromagnet, such as \ch{MnF2} or \ch{FeF2}, where the placement of the anions removes the center of inversion located between neighboring ions. For the quasi-2D AFM we take inspiration from the cuprates and consider layers of a square lattice AFM, with an additional non-magnetic ion placed in a checkboard fashion---which removes the center of inversion located between neighboring magnetic ions \cite{shanDynamicMagneticPhase2024}. Both compounds retain a global inversion symmetry. 	
	We consider a semi-infinite bulk crystal, which for the Rutile-based AFM can be treated in the three-dimensional limit, whereas the quasi-2D AFM can be treated as decoupled 2D layers. 
	
	We note in passing that these models support altermagnetism \cite{smejkalEmergingResearchLandscape2022,smejkalConventionalFerromagnetismAntiferromagnetism2022} if sufficiently long-ranged magnetic exchange paths are included \cite{liuChiralSplitMagnon2024,smejkalChiralMagnonsAltermagnetic2023, gohlkeSpuriousSymmetryEnhancement2023}. As a result, the two magnon species could exhibit a d-wave splitting \cite{smejkalChiralMagnonsAltermagnetic2023}. We neglect any discussion of altermagnetism in the following and assume degenerate magnons, since the splitting only provides corrections to the continua of magnons build from an $\alpha$ and $\beta$ magnon at finite momenta.

	\subsection{Magnon-plasmon coupling}
	We now assume that some (or all) bonds are not centers of inversion, such that the isotropic part of the polarization operator, $\bm\pi_{ij}$, is non-zero.
	We introduce the Holstein-Primakoff operators in the coupling Hamiltonian \cref{eq:Hc-pi}, to find, up to second order in Bogoliubov-transformed Holstein-Primakoff operators, the coupling Hamiltonian
	
	\begin{equation}
		H_c^{(2)}=-\ii S {\frac{1}{\sqrt{N_x N_y}}}\sum_n^{N_z}e^{- |z_n|/\lambda_{\bm q}} \sum_{\bm k\bm q}(\hat\phi_{\bm q}\bm E_{\bm q} - \hat\phi^\dagger_{-\bm q}\bm E_{-\bm q}^*)\cdot \hat{ \bm V}_{\bm k,\bm q}, \label{eq:Hc-AFM-final}
	\end{equation}
	where $\hat{\bm V}_{\bm k,\bm q}\equiv \hat{\bm V}_{\bm k,\bm q}^{\mathrm{pair}}+\hat{\bm V}_{\bm k,\bm q}^{\mathrm{scatt}}$, with
	\begin{align}
		\hat{\bm V}_{\bm k,\bm q}^{\mathrm{pair}} \equiv &u_{\bm k}u_{\bm k-\bm q} \bm\pi_{-\bm q-\bm k}^*\hat\alpha_{\bm k}\hat\beta_{-\bm k - \bm q}  \nonumber\\
		+ &u_{\bm k}u_{\bm k+\bm q} \bm\pi_{\bm q- \bm k} \hat\alpha_{\bm k}^\dagger \hat\beta_{\bm q-\bm k}^\dagger \nonumber\\ 
		+  
		&v_{\bm k}v_{\bm k-\bm q}\bm\pi_{\bm q+\bm k} \hat\alpha_{-\bm k-\bm q}\hat\beta_{\bm k} \nonumber\\
		+&v_{\bm k}v_{\bm k+\bm q}\bm\pi_{\bm k-\bm q}^* \hat\beta_{\bm k}^\dagger \hat\alpha_{\bm q-\bm k}^\dagger 
	\end{align}
	containing the pair creation and annihilation, and $\hat{\bm V}_{\bm k,\bm q}^{\mathrm{scatt}}$ describes the scattering processes, containing terms $\propto \hat\alpha_{\bm k}^\dagger\hat\alpha_{\bm k + \bm q}+\hc$ and $\propto \hat\beta_{\bm k}^\dagger\hat\beta_{\bm k + \bm q}+\hc$ Furthermore, $u_{\bm k}$ and $v_{\bm k}$ are the Bogoliubov factors defined in \cref{app:hamiltonian}.
	For simplicity, we have assumed that for every in-plane magnetic sublattice we have one atomic site in the 2DEG [such that $N_p=N_xN_y$]
	and that the plasmon electric field, $\bm E_{\bm q}(z_n)$ is slowly varying along the $z$ axis, such that the $z$-dependency can be considered as a slowly varying potential. Here the sum over $n$ is over layers at position $z_n$.
	Finally,
	\begin{equation}
		\bm \pi_{\bm k} \equiv \sum_{r_{ij}} e^{-\ii\bm r_{ij} \cdot \bm k} \bm \pi_{ij}
	\end{equation}
	is the Fourier transform of the isotropic part of the polarization tensor.

	For a Rutile AFM, the anions break the inversion symmetry along specific bonds, thus allowing for a finite polarization, as indicated for a single bond in \cref{fig:setup}. This results in the polarization
	\begin{equation}
		\bm \pi_{\bm k}^R = \ii g \cos\left(\frac{\phi_{\mathrm{nn}}^R}{2} \right)\ \begin{pmatrix}
			2c \sin(\frac12 k_x c) \sin(\frac12 k_y a) \sin(\frac12 k_z a) \\
			0.45a \cos(\frac12 k_x c) \cos(\frac12 k_y a) \sin(\frac12 k_z a)\\
			0.45a \cos(\frac12 k_x c) \sin(\frac12 k_y a) \cos(\frac12 k_z a)
		\end{pmatrix},
		\label{eq:pik-Rutiles}
	\end{equation}
	where we note that the numerical factors appear because we have assumed specific positions of the anions, taken from Ref.~\cite{hainesNeutronDiffractionStudy1997}. 
	
	For the two-dimensional AFM we have that the bond inversion symmetry is broken by the non-magnetic ions placed in a checkerboard, which gives 
	\begin{equation}
		\bm \pi_{\bm k}^{2D} = \ii g \cos\left(\frac{\phi_{\mathrm{nn}}^{2D}}{2}\right)\ \begin{pmatrix}
			a\sin( k_y a) \\
			a\sin(k_x a) \\ 
			0
		\end{pmatrix}.
		\label{eq:pik-cuprates}
	\end{equation}
	Here $g$ is a phenomenological parameter with units charge, and $\phi_{\mathrm{nn}}^{R}$, $\phi_{\mathrm{nn}}^{2D}$  is the angle the bond between magnetic ions makes with the bond from a magnetic ion to the non-magnetic ion that breaks the bond inversion symmetry. We provide more details on how to calculate $\bm \pi_{\bm k}$ in \cref{app:polarization}.
	
	Having obtained the coupling Hamiltonian, we now write down the interacting (Matsubara) plasmon Green's function \cite{stoofUltracoldQuantumFields2009,rastelliStatisticalMechanicsMagnetic2013} as
	\begin{multline}
		\mathcal G_{\bm q}(\tau) \approx \mathcal G_{\bm q}^{(0)}(\tau)\\ -\frac{1}{2!} \int_0^\beta d\tau_1 \int_0^\beta d\tau_2 \langle T_\tau H_c(\tau_1)H_c(\tau_2) \phi_{\bm q}(\tau)\phi_{\bm q}^\dagger
		\rangle_{\mathrm{con}}^{(0)}, \label{eq:Gint}
	\end{multline}
	where $ \mathcal G_{\bm q}^{(0)}(\tau)\equiv -\langle T_\tau \phi_{\bm q}(\tau)\phi_{\bm q}^\dagger \rangle^{(0)}$ is the noninteracting plasmon Green's function, $\beta\equiv(k_B T)^{-1}$ is the inverse temperature and $T_\tau$ is the time-ordering operator in imaginary time $\tau$. Finally, $\langle\dots\rangle_{\mathrm{con}}^{(0)}$ indicates averaging over the connected diagrams.
	
\begin{figure}
		\centering
		\includegraphics[width=\columnwidth]{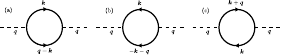}
		\caption{Feynmann diagrams contributing to the plasmon lifetime up to second order. Dashed and solid lines indicate plasmon and magnon propagators, respectively. (a) Forward, (b) backward, and (c) circle diagram. At zero temperature, only the forward and backward diagrams contribute.}
		\label{fig:feynmann}
	\end{figure}

	We can now apply Wick's theorem, to write down the resulting plasmon self-energy \cite{stoofUltracoldQuantumFields2009} up to second order in the electric field, of which there are three kinds, depicted in \cref{fig:feynmann}: (a) forward, (b) backward, and (c) circle diagrams. Since magnons are non-conserved bosonic particles, at zero temperature no magnons are present and thus the circle diagram does not contribute. We note here that it is precisely this circle diagram which produces the temperature-activated magnon-plasmon coupling in the metallic ferromagnet considered in Ref.~\cite{ghoshPlasmonmagnonInteractionsTwodimensional2023}. Since the scattering processes (described by $\hat{\bm V}_{\bm k,\bm q}^{\mathrm{scatt}}$) only enter in the circle diagram, we can neglect them from here on out. Finally, we note that the ``tadpole'' diagram does not exist, since this requires a two-plasmon-one-magnon interaction process.

	We expect the main contribution to come from the forward diagram, which gives a plasmon self-energy at zero temperature as
	\begin{equation}
		\Sigma^f_{\bm q}(\epsilon)= \frac{1}{N_xN_y}\sum_n^{N_z}e^{- |z_n|/\lambda_{\bm q}} \sum_{\bm k}\sum_{\sigma,\sigma'=\pm}\frac12 \frac{|\bm E_{\bm q}\cdot\bm{\mathcal V}_{\bm q,\bm k}^{\sigma\sigma'}|^2}{\epsilon + \ii 0^+ -\varepsilon_{\bm k} - \varepsilon_{\bm q-\bm k}}. \label{eq:self-energy}
	\end{equation}
	Here $ \bm{\mathcal V}_{\bm q,\bm k}^{\sigma\sigma'}$ is the interaction vertex in the Boguliobov basis, given by 
	\begin{equation}
		\bm{\mathcal V}_{\bm q,\bm k}\equiv S
		\begin{pmatrix}
			0 & u_{\bm k} u_{\bm k +\bm q} \bm \pi_{\bm q-\bm k} \\
			v_{\bm k} v_{\bm k +\bm q} \bm \pi_{-\bm q+\bm k}& 0
		\end{pmatrix}.
	\end{equation}
	We note that $u_{\bm k}u_{\bm k+\bm q} $ and $v_{\bm k}v_{\bm k+ \bm q} $ reduce to the usual exchange enhancement factors for $q\ll k$ \cite{dyrdalMagnonplasmonHybridizationMediated2023}, which is typically the case given the mismatch in wavelength between magnons and plasmons. The backward diagram gives 
	\begin{equation}
		\Sigma^b_{\bm q}(\epsilon)= \frac{-1}{N_xN_y}\sum_n^{N_z}e^{- |z_n|/\lambda_{\bm q}} \sum_{\bm k}\sum_{\sigma,\sigma'=\pm}\frac12 \frac{|\bm E_{\bm q}\cdot\bm{\mathcal V}_{ -\bm q,\bm k}^{\sigma\sigma'}|^2}{\epsilon  +\varepsilon_{\bm k} + \varepsilon_{\bm q-\bm k}}, \label{eq:sigma-backward}
	\end{equation} 
	which is purely real. The combined plasmon self-energy is now defined as
	\begin{equation}
		\Sigma_{\bm q}(\epsilon) \equiv \Sigma^f_{\bm q}(\epsilon) + \Sigma^b_{\bm q}(\epsilon).
	\end{equation}
	In addition, we expect that the self-energy following from the backward diagram [\cref{eq:sigma-backward}] is suppressed by the magnon bandwidth, given the $\varepsilon_{\bm k} + \varepsilon_{\bm q-\bm k}$ terms in the denominator. We have numerically confirmed that typically $\Sigma_{\bm q}^b(\epsilon)\ll|\Sigma_{\bm q}^f(\epsilon)|$, and we thus discuss only the forward diagram from here on. In the numerical result that follow, we do however include the backward diagram for completeness. The circle diagram is strictly zero at zero temperature.

	The self energy in \cref{eq:self-energy} is valid in two and three dimensions, and takes into account the decay of the plasmon electric field. We now assume that the antiferromagnet is much thicker than the plasmon decay length, such that we can take $N_z\rightarrow\infty$ and explicitly  perform the summation over $n$ to find $\sum_n^{\infty}e^{- |z_n|/\lambda_{\bm q}}=(e^{2a_z/\lambda_{\bm q}}-1)^{-1}$, where $a_z$ is the lattice constant along the $z$ axis.\footnote{For a thin film thinner than the decay length, one instead obtains $(1-e^{-2a_zN_z/\lambda_{\bm q}})/(e^{2a_z/\lambda_{\bm q}}-1)$. } This scaling corresponds to the fact that a large number of magnon modes are available for a plasmon mode to decay into, which is only limited by the decay length of the plasmon electric field. This is similar to how magnon-photon coupling in a cavity is is proportional to the square root of the number of exchange-coupled spins \cite{hueblHighCooperativityCoupled2013}.

	We furthermore define here the spin-zero two-magnon density of states (DOS) as  
	\begin{equation}
		D_{\bm q}(\epsilon)= \frac{1}{N_m}\sum_{\bm k} \delta({\epsilon -\varepsilon_{\bm k} - \varepsilon_{\bm q-\bm k}}) \label{eq:2DOS},
	\end{equation}
	which we refer to as ``spin-zero'' because no change in net spin is associated with the pair creation of magnons with momenta $\bm k,\bm q-\bm k$. In other words, it corresponds to the simultaneous pair creation of magnons, $\sim\alpha_{\bm k}^\dagger\beta_{\bm q-\bm k}^\dagger $ and $\sim\beta_{\bm k}^\dagger\alpha_{\bm q-\bm k}^\dagger $, which in an AFM does not require spin-orbit coupling and thus conserves spin. It is therefore directly related to the imaginary part of the plasmon self-energy,
	\begin{equation}
		\Im[\Sigma_{\bm q}(\epsilon)] \propto \frac{1}{N_x N_y}\sum_{\bm k}\sum_{\sigma,\sigma'=\pm} {|\bm E_{\bm q}\cdot\bm{\mathcal V}_{\bm k,\bm q}^{\sigma\sigma'}|^2}\delta({\epsilon -\varepsilon_{\bm k} - \varepsilon_{\bm q-\bm k}}), \label{eq:ImEpsilon}
	\end{equation}
	weighed by the interaction vertex.  
	The spin-zero two-magnon DOS therefore plays an important role in determining the plasmon self-energy. 
	
	For the antiferromagnetic models considered here, the edges of the spin-zero two-magnon DOS are given by
	 $2\varepsilon_{\mathrm{min}}$ and $2\varepsilon_{\mathrm{max}}$, where
	 \begin{align}
	 	\varepsilon_{\mathrm{min}}&=\sqrt{K(K+2zJ)} \\
	 	\varepsilon_\mathrm{max}&=K+zJ,
	 \end{align}
	 with $z$ the number of nearest neighbors. Importantly, at the upper edge of the continuum a Van Hove singularity in the spin-zero two-magnon DOS is located, which is related to the pair creation of magnons on the Brillouin zone boundary. We show the spin-zero two-magnon DOS in \subfigref{fig:rutile}{(a)} and \subfigref{fig:cuprate}{(a)} for the Rutile and quasi-2D AFM respectively.

	It is important to take into account that higher-order magnon-magnon interactions will renormalize the plasmon self-energy, since the two excited magnons  are close to each other in real space and will thus interact through magnon-magnon interactions. The main effect of this interaction is to cut the Van Hove singularity at the Brillouin zone edge. We take this broadening into account by introducing a finite magnon lifetime, $\varepsilon_{\bm k} \rightarrow \varepsilon_{\bm k}-\ii \kappa$, where $\kappa=0.1\varepsilon_{\mathrm{max}}$ is chosen to reproduce the broadening calculated through many-body theory in the context of Raman scattering \cite{canaliTheoryRamanScattering1992}. The magnon-magnon interaction will also induce a small energy shift, which we neglect here.
	
	The plasmon self-energy is derived perturbatively, and thus assumes that the magnon-plasmon coupling is small. This is however not a straightforward requirement, since the plasmon self-energy is strongly enhanced due to the Van Hove singularity at the zone boundary. We therefore additionally require that the plasmon self-energy is smaller than the plasmon energy, i.e., $|\Sigma_{\bm q}(\epsilon)|<\epsilon$. We enforce this requirement by choosing a suitably small coupling constant $g$. Beyond the perturbative limit, i.e., if $|\Sigma_{\bm q}(\epsilon)|>\epsilon$, this system is in the deep-strong coupling regime \cite{friskkockumUltrastrongCouplingLight2019}. Here, the perturbative expansion we perform breaks down, and a many-body approach is needed \cite{verresenAvoidedQuasiparticleDecay2019}.

	\section{Results}
	\label{sec:results}
	We now analyze the renormalized plasmon spectrum, by calculating the spectral function \cite{stoofUltracoldQuantumFields2009}
	\begin{equation}
		A_{\bm q}(\epsilon) = -\frac{1}{\pi}\Im[\mathcal G_{\bm q}(\epsilon)],
	\end{equation}
	where 
	\begin{equation}
		\mathcal G_{\bm q}(\epsilon) = \left[\epsilon + \ii 0^+ - \mathcal{E}_{\bm q}(1-\ii\gamma) - \Sigma_{\bm q}(\epsilon) \right]^{-1}
	\end{equation}
	is the plasmon retarded Greens function. Here $\mathcal E_{\bm q}\propto\sqrt{q}$ is the plasmon dispersion, which we give in \cref{app:plasmon}. We have included a finite plasmon lifetime through the inverse plasmon quality factor, which we set to $\gamma=\gammaplasmon$ \cite{principiIntrinsicLifetimeDirac2013}. 
	Furthermore, we set $g=\gpolarizationrutile\times e$, which is comparable in magnitude to the polarization amplitude calculated for \ch{Sr2IrO4} \cite{seifertOpticalExcitationMagnons2019} and \ch{Cr2O3} \cite{mostovoyTemperatureDependentMagnetoelectricEffect2010} and for a Hubbard model with $t/U=0.1$ \cite{zhuSpinorbitCouplingElectronic2014}. These polarization amplitudes were chosen to remain in the perturbative regime, i.e., such that we have $|\Sigma_{\bm q}(\epsilon)|<\epsilon$. For simplicity, we also consider the Rutile-AFM to  be cubic, i.e., $a=c$. Then we have $\phi_{\mathrm{nn}}^{R}=\SI{119}{\degree}$, whereas for the quasi-2D AFM we have $\phi_{\mathrm{nn}}^{2D}=\SI{90}{\degree}$.
	
	\subsection{Rutile-based antiferromagnet}
	\begin{figure*}
		\includegraphics{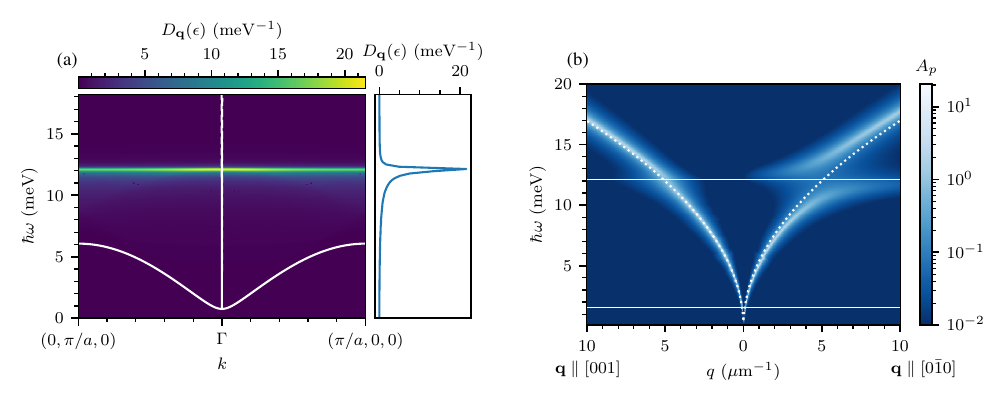}
		\caption{Coupling between plasmons and the magnon continuum in rutile-type antiferromagnets [\subfigref{fig:setup}{b}]. (a) Magnon dispersion in the $k_x-k_y$ plane, with the white line indicating $\varepsilon_{\bm{k}}$. The dashed white line indicates the steep plasmon dispersion and the colorscale is the spin-zero two-magnon density of states. Shown on the right is a cut of the spin-zero two-magnon density of states along the plasmon energy, i.e., $D_{\bm q}(\mathcal E_{\bm q})$.
			(b) The plasmon spectral function $A_{\bm q}(\epsilon)$ along two propagation directions of the plasmon: $\bm q \parallel [0\bar10] \parallel \hat x$ and $\bm q \parallel [001]\parallel \hat y$ (cf.~Fig.~\ref{fig:setup} for a real space geometry). The lower and upper edges of the spin-zero two-magnon continuum are indicated by the horizontal white lines. The bare plasmon dispersion is indicated by the white dotted line. We stress that the plasmon spectral function is shown on a logarithmic scale. \label{fig:rutile}}
	\end{figure*}

	We show the resulting plasmon spectral function for a Rutile-based AFM in \subfigref{fig:rutile}{(b)}, as a function of wave vector $\bm q$ and energy $\epsilon$, for a plasmon propagating along ${x}$ and ${y}$. We have chosen the wave vector range such that the plasmon energies are comparable to the magnon energies. The white dotted line indicates the bare plasmon dispersion. 
	The lower and upper edges of the spin-zero two-magnon continuum are indicated by the horizontal white lines. We remind the reader that the spin-zero two-magnon continuum is directly related to the imaginary part of the plasmon self-energy [\cref{eq:ImEpsilon}], and is thus related to the plasmon decay.
	For the range of wave vectors considered here, the decay length $\lambda$ is comparable to the wave length, i.e., of the order of several micrometers, which will significantly enhance the coupling.

	We first focus on a plasmon propagating along ${x}$, where we observe that below the spin-zero two-magnon continuum the plasmon dispersion is unaffected. This is to be expected, since here the imaginary part of the plasmon self-energy is strictly zero. Upon entering the spin-zero two-magnon continuum, the plasmon decays into the spin-zero two-magnon continuum \cite{verresenAvoidedQuasiparticleDecay2019}, in particular at the Van Hove singularity, which is located at the upper edge of the continuum.
	After exiting the spin-zero two-magnon continuum, the plasmon becomes visible again, but its dispersion is normalized---as can be seen by comparing the plasmon spectral function above the spin-zero two-magnon continuum with the bare dispersion (white dotted line). 

	We note here that the plasmons also generate a magnetic field, which couples to the magnons through the Zeeman coupling \cite{costaStronglyCoupledMagnonplasmon2023,yuanStrongTunableCoupling2024,yuanBreakingSurfacePlasmonExcitation2024}. This process gives a bilinear plasmon-magnon coupling, as opposed to the quadratic in magnon operators coupling we consider here. This coupling therefore gives an avoided crossing where the plasmon crosses the magnon branch, which (given the square root dispersion of the plasmon) happens at the magnon gap. This crossing can therefore be readily distinguished from the coupling to the spin-zero two-magnon continuum, which happens at much higher energies.

	At this crossing, we have for the current parameter set that $|\Sigma_{\bm q}(\epsilon)|/\epsilon\approx0.4$, highlighting that we are in the ultrastrong coupling regime. Here we define ultrastrong coupling as the regime where the coupling strength is a significant fraction of the bare resonance frequency \cite{friskkockumUltrastrongCouplingLight2019}.
	Furthermore, since the polarization strength $g$ enters the coupling quadratically [cf. \cref{eq:self-energy}], it is also relatively easy to reach the deep-strong coupling regime with  slightly stronger polarization strengths. We define the deep-strong coupling regime as the regime where the coupling strength exceeds the bare resonance frequency and thus $|\Sigma_{\bm q}(\epsilon)|>\epsilon$.
	 As was explained before, in the deep-strong coupling regime our perturbative treatment breaks down and we are unable to reliably calculate the full plasmon spectral function. In the deep-strong coupling  regime, it might also be possible to observe the expulsion of the plasmon from the Van Hove singularity \cite{verresenAvoidedQuasiparticleDecay2019}.
	
	We next turn to a plasmon along ${y}$, where the coupling to the continuum is much weaker, and there is no noticeable decay. 
	This can be explained by the electric field of the plasmon, which is parallel to its propagation direction, in addition to an out-of-plane electric field, i.e., $\bm E =  E_{\parallel}\hat{\bm q} + E_z \hat{\bm z}$ [\cref{app:plasmon}]. Thus, a plasmon along ${y}$ only couples to the $y$ and $z$ component of the polarization vector, $\bm\pi_{\bm k}$ [\cref{eq:pik-Rutiles}]. Because of the Van Hove singularity at the Brillouin zone boundaries, we furthermore have that mainly magnons at the Brillouin zone boundaries contribute to the plasmon self-energy. However, the $y$-component of the polarization vanishes at eight of the twelve edges of the Brillouin zone, whereas the $x$-component of the polarization has a maximum on all edges of the Brillouin zone, and the $z$-component of the polarization has a maximum only along the $k_y=\pm \pi/a$ edges of the Brillouin zone. A plasmon propagating along $x$ therefore couples much stronger to the spin-zero two-magnon continuum. We note that the symmetries of $\bm\pi_{\bm k}$ are set by the crystal symmetry, and could thus change in different compounds---offering a potential way to change the strength of the magnon-plasmon coupling. In particular, for the Rutiles grown along the [001]-axis, $\bm\pi_{\bm k}$ inherits the $C_4^z$ point group symmetry of the lattice and thus the self energy is equivalent along $x$ and $y$.
	
	The imaginary part of the plasmon self-energy is only non-zero if the spin-zero two-magnon density of states, \cref{eq:2DOS}, is finite. To explore this in more detail, we show the spin-zero two-magnon continuum in \subfigref{fig:rutile}{(a)}. We also show the spin-zero two-magnon DOS for the on-shell plasmon energies in the \subfigref{fig:rutile}{(b)}. 
	We first note that due to the wavelength mismatch between magnons and plasmons in a Rutile AFM, the plasmons have effectively zero wavevector compared to the magnons.  The two-magnon DOS contains two analytic discontinuities: upon entering the continuum, at $\epsilon=2\sqrt{K(K+2zJ)}$, and upon exiting the continuum at $\epsilon=2K+2zJ$. In three dimensions these discontinuities produce a square root behavior of the real part of the plasmon self-energy \cite{chernyshevSpinWavesTriangular2009,zhitomirskyColloquiumSpontaneousMagnon2013}. Secondly, the spin-zero two-magnon DOS peaks strongly at the upper edge of the spin-zero two-magnon continuum, where a Von Hove singularity is located, corresponding to the excitation of two zone edge boundaries, where the magnon dispersion is flat. 
	These two features of the spin-zero two-magnon DOS explain why the spin-zero two-magnon continuum repels the plasmon, and why this repelling originates from the upper edge of the spin-zero two-magnon continuum. Note that the Van Hove singularity is only a true singularity for noninteracting magnons, but is cut by the magnon lifetimes \cite{bayrakciLifetimesAntiferromagneticMagnons2013} and higher order magnon interactions \cite{canaliTheoryRamanScattering1992}, which we have modeled here by introducing a finite magnon lifetime. We also stress that at the Brillouin zone edge, the exchange enhancement factors tend to unity, $u_{\bm k}^2\approx v_{\bm k}^2 \approx1$.  In the plasmon self-energy we thus do not see a significant exchange enhancement.

	\subsection{Quasi-two dimensional antiferromagnet}
	
\begin{figure*}
		 \includegraphics{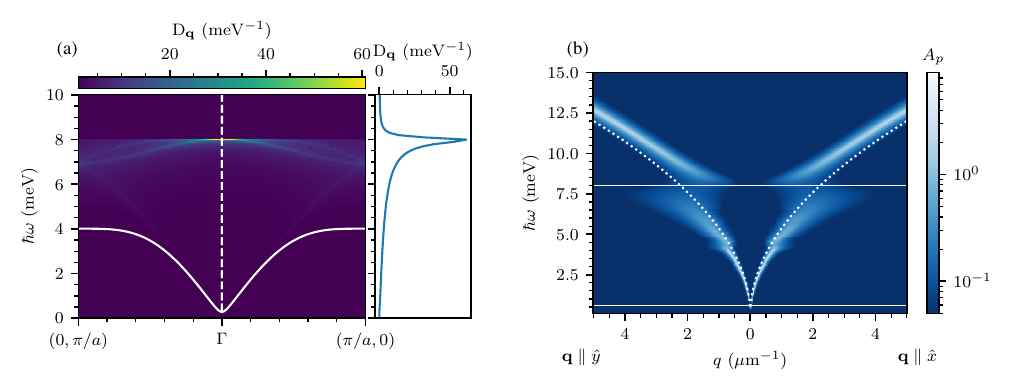}
		\caption{Coupling between plasmons and the magnon continuum in a quasi two-dimensional antiferromagnet [\subfigref{fig:setup}{c}]. (a) Magnon dispersion in the $k_x-k_y$ plane, with the white line indicating $\varepsilon_{\bm{k}}$. The dashed white line indicates the steep plasmon dispersion and the colorscale is the spin-zero two-magnon density of states. Shown on the right is a cut of the spin-zero two-magnon density of states along the plasmon energy, i.e., $D_{\bm q}(\mathcal E_{\bm q})$. 
			(b) The plasmon spectral function $A_{\bm q}(\epsilon)$ along two propagation directions of the plasmon: $\bm q \parallel \hat{x}$ and $\bm q \parallel \hat{y}$ [see Fig.~\ref{fig:setup}(a) for real space geometry]. The lower and upper edges of the spin-zero two-magnon continuum are indicated by the horizontal white lines. The bare plasmon dispersion is indicated by the white dotted line. We stress that the plasmon spectral function is shown on a logarithmic scale. \label{fig:cuprate}}
	\end{figure*}

	We now consider the quasi-2D AFM, and show the resulting plasmon spectral function in \cref{fig:cuprate}, considering the propagation of plasmons along $ x$ and $ y$. As before, the white dotted line indicates the bare plasmon dispersion and the lower and upper edges of the spin-zero two-magnon continuum are indicated by the horizontal white lines.

	From the plasmon spectral function in \subfigref{fig:cuprate}{(b)} we observe that the plasmon again strongly decays as its frequency matches the Van Hove singularity in the spin-zero two-magnon DOS, similar to the results obtained for the Rutile AFM. For plasmon frequencies above the continuum, the plasmon is again renormalized. This therefore qualitatively reproduces the results for the three-dimensional Rutile-based AFM. Note again that we have restricted ourselves to remain in the ultrastrong coupling regime, but the deep-strong coupling regime can again be easily reached with a slightly higher polarization strength. We also stress here that it is difficult to make direct comparisons between the rutile and quasi-2D AFM, since they are different models.
	
	The polarization is invariant under rotation in the plane, and thus the plasmon self-energy is identical for all propagation directions---as can also be seen by comparing the propagation along $x$ and $y$ in \subfigref{fig:cuprate}{(b)}.

	Finally, we note that in two dimensions the discontinuities in the imaginary  part of the plasmon self-energy upon entering and exciting the spin-zero two-magnon continuum are step functions, which in two dimensions implies a logarithmic divergence in the real part \cite{chernyshevSpinWavesTriangular2009,zhitomirskyColloquiumSpontaneousMagnon2013}. This could potentially enhance the magnon-plasmon interaction compared to the three-dimensional Rutile AFM, but this divergence is smeared out by the magnon-magnon interactions \cite{canaliTheoryRamanScattering1992}. Furthermore, at the lower edge of the spin-zero two-magnon continuum, the continuum is formed of two magnons with (almost) zero momentum, for which the polarization, \cref{eq:pik-cuprates}, vanishes, and thus the plasmon self-energy is insignificant---which we confirmed numerically (not shown here).

	\section{Conclusion and discussion}
	\label{sec:conclusion}
	We have shown in this work that plasmons couple to the spin-zero two-magnon continuum in antiferromagnetic thin films with a finite bond polarization. We find that the plasmon mode decays as it enters the Van Hove singularity at the upper edge of the spin-zero two-magnon continuum, and is renormalized upon exiting the continuum again. Furthermore, we predict that the ultrastrong and potentially deep-strong coupling regime could be reached in Rutile and quasi-two-dimensional antiferromagnets---providing a new platform to study ultrastrong light-matter coupling.
	
	One could reasonably ask the question if the large plasmon self-energies are not due to a failure of the perturbation theory used to obtain the self-energy expressions. Firstly, we note that we have neglected here the magnon self-energies which also arise from the same coupling Hamiltonian \cref{eq:Hc-AFM-final}. We show however in \cref{app:self-energies-magnon} that the decay of a magnon into a plasmon and a magnon is kinematically not allowed, due to the strong mismatch in wavelength between the magnons and plasmons---and thus the magnon self-energy vanishes. Up to second order in the plasmon electric field $\bm E_{\bm q}$ our theory is therefore complete. This leaves diagrams which are of higher order in $\bm E_{\bm q}$ to potentially renormalize the coupling of the plasmon to the spin-zero two-magnon continuum. However, since $\bm E_{\bm q}$ is typically small, we do not expect this effect to be significant. This leaves the magneto-optical couplings which are typically considered in two-magnon Raman scattering, which are quadratic in plasmon operators. We show in \cref{app:magneto-optic} that these self energies have a reduced phase space available and we thus expect this self energy to be significantly weaker. We note however that the Raman scattering will be the leading contribution if all bonds are inversion symmetric, since in that case the polarization operator $\bm \pi$ vanishes.

	As a first step, we have considered a heterostructure of a 2DEG and an insulating AFM, such that the magnons and plasmon are separated in space and couple through the plasmons electric field. We also expect that within a metallic antiferromagnet, which could host both plasmons and magnons, the same coupling could occur, as was already shown by Ref.~\cite{ghoshPlasmonmagnonInteractionsTwodimensional2023} for metallic ferromagnets. Here the coupling would potentially be enlarged, since the plasmons and magnons are no longer physically separated.
	
	Within the AFMs models considered in this work, the polarization strength, $g$, is a critical parameter to strongly couple the plasmon to the spin-zero two-magnon continuum. We note here that several antiferromatic materials with a large polarization strength exist, such as multi-orbital magnetic Mott insulators \cite{bolensTheoryElectronicMagnetoelectric2018}, \ch{Sr2IrO4} \cite{seifertOpticalExcitationMagnons2019} and \ch{Cr2O3} \cite{mostovoyTemperatureDependentMagnetoelectricEffect2010}. Additionally, cuprates are a good candidate \cite{curtisCavityMagnonpolaritonsCuprate2022,shanDynamicMagneticPhase2024}, and are typically well-described by the Fermi-Hubbard model, which allows one to calculate the polarization \cite{zhuSpinorbitCouplingElectronic2014}. However, we note here that the large magnon bandwidth (in the order of 100s of meV) might prohibit the coupling of plasmons to the spin-zero two-magnon continuum, since typical plasmons energies do not extend to such large values. Alternatively, it could be possible to engineer higher-energy plasmons, such as by making use of interband plasmons in moir\'e graphene-metal heterostructures \cite{politanoEffectMoireSuperlattice2017}. Another exciting future direction is the coupling of plasmons to multiferroics, where the magnetoelectric effects can be large \cite{kimuraMagneticControlFerroelectric2003, wangEpitaxialBiFeO3Multiferroic2003}, and the electrically active magnons \cite{katsuraDynamicalMagnetoelectricCoupling2007, pimenovPossibleEvidenceElectromagnons2006, sushkovElectromagnonsMultiferroicYMn2O52007} may enhance the coupling to plasmons.

	\begin{acknowledgments}
		It is a pleasure to thank Davide Bossini for stimulating discussions. This work is in part funded by the Deutsche Forschungsgemeinschaft (DFG, German Research Foundation) -- Project No.~504261060 (Emmy Noether Programme). P.~G. acknowledges financial support from the Alexander von Humboldt postdoctoral fellowship.
	\end{acknowledgments}
	The data that support the findings of this article are openly available \cite{pgunninkPgunninkMagnonplasmonlifetimeV12025}.

	\appendix
	
	\section{Plasmon dispersion and generated electric field}
	\label{app:plasmon}
	Here we reproduce for completeness the dispersion and generated electric field of plasmons, following \textcite{ferreiraQuantizationGraphenePlasmons2020}. We consider a 2DEG surrounded by a dielectric medium with dielectric constant $\epsilon$. This gives the plasmon Hamiltonian as
	\begin{equation}
		H_p = \sum_{\bm q} \omega_{\bm q,p} \hat \psi_{\bm{q}}^\dagger \hat\psi_{\bm q},
	\end{equation} where plasmon dispersion is defined through
	\begin{equation}
		\omega_{\bm q,p}^2 = \frac{1}{2}\frac{D}{2\epsilon_0 \epsilon}\sqrt{4q^2 + \left(\frac{D}{2\epsilon_0 c^2}\right)^2} - \frac{1}{2}\left(\frac{D}{2\epsilon_0 \sqrt{\epsilon}c}\right)^2.
	\end{equation}
	Here, $D$ is the Drude weight, which for simplicity we take to be $D = e^2 E_F / (\pi\hbar^2)$, as is the case for graphene. The electric field can then be found as
	\begin{equation}
		\bm E(\bm r) = \ii\sqrt{\frac{\hbar\omega_{\bm q,p} }{2A\epsilon_0 N_{\bm q}}} e^{\ii\bm q \cdot \bm r} \left(i \hat{\bm {q}} - \frac{q}{\kappa_{\bm q}}\sign{z}\hat{\bm z} \right) e^{- z/\lambda_{\bm q}},
	\end{equation}
	where 
	\begin{equation}
		N_{\bm q} = \frac{\epsilon}{\kappa^3_{\bm q}}\left(\kappa_{\bm q}^2 + q^2\right)
	\end{equation}
	is the mode length and
	\begin{equation}
		\kappa_{\bm q} = \sqrt{q^2 - \omega_{\bm q,p}^2\epsilon/c^2}
	\end{equation}
	is the out-of-plane momentum.  Finally, $\lambda_{\bm q}=\kappa_{\bm q}^{-1}$ is the decay length along the $z$-axis.
	Throughout this work, we take $E_F=\Efermi$ and a typical  dielectric constant for insulating compounds, $\epsilon=\epsilonplasmon$, which holds for example for \ch{MnF2}  \cite{seehraEffectTemperatureAntiferromagnetic1986}.

	\section{Antiferromagnetic magnon Hamiltonian}
	
	\label{app:hamiltonian}
	
	We introduce the Holstein-Primakoff operators 
	\begin{align}
		\hat S_{A;i}^+ &\approx \sqrt{2S} \hat a_i; \quad &\hat S_{A;i}^z&= S- \hat a_i^\dagger \hat a_i; \\
		\hat S_{B;i}^+ &\approx \sqrt{2S} \hat b_i^\dagger; \quad &\hat S_{B;i}^z&= -S+ \hat b_i^\dagger \hat b_i 
	\end{align}
	in the Heisenberg Hamiltonian \cref{eq:HAFM} and apply the Fourier transformation, $\hat a_i = \frac{1}{\sqrt{N_m}}\sum_{\bm k} e^{i\bm k\cdot \bm r_i}\hat a_{\bm k}$, where $N_m = N_x \times N_y \times N_z$ is the number of spins in each sublattice, to find the Hamiltonian for $\eta\in\{\mathrm{Rutile},{2D}\}$,
	\begin{multline}
		\hat H_{\mathrm{AFM;\eta}}^{(2)}=\sum_{\bm k} (zJ+K)\hat a_{\bm k}^\dagger \hat a_{\bm k} +(zJ+K)\hat b_{\bm k}^\dagger \hat b_{\bm k}\\+J_{\eta;\bm k}(\hat a_{\bm k}\hat b_{-\bm k} + \hat a_{\bm k}^\dagger \hat b_{-\bm k}^\dagger)
	\end{multline} 
	with
	\begin{align}
		J_{\mathrm{Rutile};\bm k} &= zJ\cos (\frac12k_x a)\cos (\frac12k_y a)\cos (\frac12k_z a), \\ 
		J_{{2D};\bm k} &= \frac12zJ\left[\cos ( k_x a) + \cos (k_y a)\right],
	\end{align}
	where $z$ is the number of nearest neighbhors, i.e., $z=8$ for the Rutile-like AFM and $z=4$ for the quasi-2D AFM.
	This Hamiltonian can be diagonalized through the Bogoliubov transformation, 
	\begin{align}
		\hat a_{\bm k}& = u_{\eta;\bm k} \hat \alpha_{\eta;\bm k} - v_{\eta;\bm k}  \hat\beta_{-\bm k}^\dagger \nonumber \\ 
		\hat b^\dagger_{-\bm k} &= -v_{\eta;\bm k} \hat \alpha_{\bm k} + u_{\eta;\bm k}  \hat\beta_{-\bm k}^\dagger \label{eq:bog-transformation}
	\end{align} 
	to find, up to a constant,
	\begin{equation}
		\hat H_{\eta}=\sum_{\bm k} \epsilon_{\eta;\bm k} ( \hat \alpha_{\bm k}^\dagger \hat \alpha_{\bm k} + \hat\beta_{\bm k}^\dagger \hat\beta_{\bm k} ),
	\end{equation}
	with
	\begin{equation}
		\epsilon_{\eta;\bm k} =  \sqrt{(zJ_\eta+K)^2-J_{\eta;\bm k}^2}\label{eq:dispersion-afm-square}
	\end{equation}
	as the antiferromagnetic dispersion relation and 
	\begin{equation}
		u_{\eta;\bm k} = \cosh\varphi_{\eta;\bm k};\quad v_{\eta;\bm k}=\sinh\varphi_{\eta;\bm k} \label{eq:bog-factors}
	\end{equation}
	the Boguliobov factors, where
	\begin{equation}
		\tanh(2\varphi_{\eta;\bm k}) \equiv \frac{-J_{\eta;\bm k}}{zJ_\eta + K + J_{\eta;\bm k}'}.
	\end{equation}

	Throughout this work, we set $S=1$. For the Rutile AFM, $J=\Jexchange$, $K=\Kani$ (parameters inspired by \ch{MnF2} \cite{rezendeIntroductionAntiferromagneticMagnons2019}) and for the quasi-2D AFM, we take $J=\Jexchangecupr$, $K=\Kanicupr$. Note that in the Rutile-AFM, the anions could introduce a non-zero further-neighbor coupling, which leads to an altermagnetic splitting of the band structure \cite{smejkalChiralMagnonsAltermagnetic2023,gohlkeSpuriousSymmetryEnhancement2023}, but this is splitting has negligible effects on the magnon continuum close to the $\Gamma$ point. We have therefore neglected it here. Similarly, the non-magnetic ions in the quasi-two dimensional AFM could introduce a non-zero further-neighbor coupling and thus an altermagnetic spin splitting \cite{brekkeTwodimensionalAltermagnetsSuperconductivity2023}.
	
	\section{Polarization operator}
	\label{app:polarization}
	In absence of spin-orbit coupling, the polarization operator $\bm \pi_{ij}$ for the bond connecting sites $i$ and $j$ is given by \cite{zhuSpinorbitCouplingElectronic2014}
	\begin{equation}
		\bm \pi_{ij} = g_{ij} a_{ij}(\bm e_{jk} - \bm e_{ki}),
	\end{equation}
	where $k$ is any additional site which mediates hopping between sites $i$ and $j$, and summation over $k$ is implied if there are multiple sites mediating hopping between $i$ and $j$. $\bm e_{ij}$ is the unit vector connecting sites $i$ and $j$, $a_{ij}$ is the distance between sites $i$ and $j$ and $g_{ij}$ is a phenomenological parameter describing the polarization amplitude, with units charge. 
	
	\begin{figure}
		\centering
		\includegraphics[width=0.6\columnwidth]{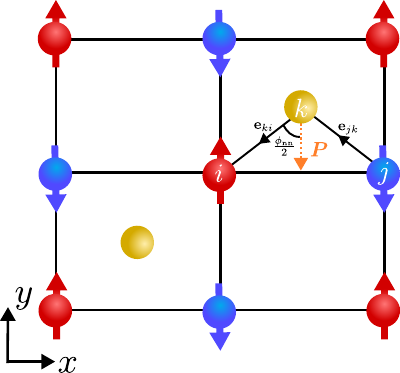}
		\caption{For the quasi-2D AFM, the polarization operator illustrated between two magnetic sites $i$ and $j$, as mediated by a non-magnetic site $k$. \label{fig:angles}}
	\end{figure}
	
	We illustrate this by working out the polarization operator for the quasi-2D AFM, as illustrated in \cref{fig:angles}. For the two sites $i$ and $j$ indicated, the 
	two relevant unit vectors are $\bm e_{ki} = \tfrac{1}{\sqrt{2}}(-1,-1)$ and $\bm e_{jk}=\tfrac{1}{\sqrt{2}}(-1,1)$, such that the direction of the polarization operator is $(\bm e_{jk}-\bm e_{ki})=(0,-\sqrt{2})$, as indicated. Alternatively, we can directly write down $(\bm e_{jk}-\bm e_{ki})=\frac12\cos(\phi_{\mathrm{nn}}/2)$, where $\phi_{\mathrm{nn}}$ is the angle of the bond. The latter form is what we use \cref{eq:pik-Rutiles,eq:pik-cuprates}. We then directly obtain that
	\begin{equation}
		\bm \pi_{ij} = \frac{1}{2}ag\cos\left(\frac{\phi_{\mathrm{nn}}}{2}\right)\ \hat{y}
	\end{equation}
	for the two sites indicated in \cref{fig:angles}. After performing the summation over all four neighboring sites and performing the Fourier transform, we obtain \cref{eq:pik-cuprates}.
	
	\section{Full self-energy diagrams}
	\label{app:self-energies-magnon}
\begin{figure}
		\centering
		\includegraphics[width=\columnwidth]{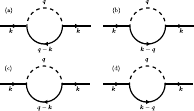}
		\caption{Feynmann diagrams contributing to the magnon lifetime up to second order.  Dashed and solid lines indicate plasmon and magnon propagators respectively. (a) and (b) circle diagrams, (c) backward diagram, (d) forward diagram. At zero temperature only the forward diagram remains.  }
		\label{fig:feynmann-magnons}
	\end{figure}

	In addition to the two plasmon self-energy diagrams in \cref{fig:feynmann}, the magnons will also gain a self energy represented by the diagrams in \cref{fig:feynmann-magnons}. To be fully consistent, these magnon self-energies would have to be solved in tandem with the plasmon self-energies \cite{chernyshevSpinWavesTriangular2009}. In particular, a finite imaginary part of the magnon self-energy would cut-off the singularity at the upper edge of the spin-zero two-magnon continuum. Note that the magnons already have a finite lifetime, due to relaxation processes and higher-order magnon-magnon interactions, and thus the singularity is already partly cut.
	
	We will now analyze the possible magnon self energies in \cref{fig:feynmann-magnons}. For simplicity, we assume a finite thickness $N_z$ and a constant plasmon electric field, i.e., $N_za_z\ll\lambda_{\bm q}$.
	We first note that the circle diagrams (a-b) vanish at zero temperature. The backward diagram (c) will only introduce a real self-energy and we are thus left with the forward diagram in (d), which is given by  
	\begin{equation}
		\Sigma(\bm k, \epsilon) = \frac{1}{N_p}\sum_{\bm q} \frac12 \frac{|\bm E_{\bm q}\cdot\bm{\mathcal W}_{\bm q,\bm k}|^2}{\epsilon - \epsilon_{\bm k - \bm q} - \mathcal E_{\bm q} + \ii 0^+},
	\end{equation}
	where $\bm{\mathcal W}_{\bm q,\bm k}^{\sigma}$ is an interaction vertex, which can be found from \cref{eq:Hc-AFM-final}. As a first step, we assume the magnon energy to be on-shell, i.e., $\epsilon=\epsilon_{\bm k}$, such that we can write the imaginary part of this self energy as
	\begin{equation}
		\Im\Sigma(\bm k) = -\frac{\pi}{N_p}\sum_{\bm q} \frac12 |\bm E_{\bm q}\cdot\bm{\mathcal W}_{\bm q,\bm k}^{\sigma}|^2\delta(\epsilon_{\bm k} - \epsilon_{\bm k - \bm q} - \mathcal E_{\bm q}).
	\end{equation}
	Since we have that $q\ll k$, we expand the magnon energies in small $q$ to find $\epsilon_{\bm k - \bm q}\approx \epsilon_{\bm k} - \bm q\cdot\bm v_{\bm k}$, where $\bm v_{\bm k}\equiv \nabla_{\bm k}\epsilon_{\bm k}$ is the group velocity. The delta function then becomes $		\delta(\bm q\cdot\bm v_{\bm k} - \mathcal E_{\bm q})$, implying that in order for the imaginary part of the above self energy to be non-zero at any part of the magnon Brillouin zone, one requires
	\begin{equation}
		\max_{\mathrm{BZ}}|v_{\bm k}| > \mathcal E_{\bm q} / q.
	\end{equation}

	However, given the steep plasmon dispersion compared to the magnon dispersion, this is not possible for realistic parameters.\footnote{In particular, note that $\mathcal E_{\bm q}\propto q^{1/2}$ and thus $\mathcal E_{\bm q}/q\propto q^{-1/2}$.} We thus conclude that there is no imaginary part of the above self energy which could renormalize the singularity at the upper edge of the spin-zero two-magnon continuum.
	
	\section{Plasmon self-energy following from the magneto-optic interaction }
	\label{app:magneto-optic}
	Here we consider the quadratic magneto-optic interaction, which for simple cubic lattices gives the off-resonant interaction \cite{fleuryScatteringLightOne1968,parviniCavityenhancedOpticalManipulation2025} 
	\begin{equation}
		\hat H_{q} = -\sum_{ij} B( \hat{\bm E}_i\cdot \bm r_{ij})( \hat{\bm E}_j^\dagger\cdot \bm r_{ij}) \hat{\bm S}_{i}\cdot\hat{\bm S}_j.
	\end{equation}
	Here $B$ is a phenomological coupling strength. In a half-filled single-band Hubbard model with hopping $t$ and on-site interaction $U$ it is given by $B={4t^2}/{(U-\hbar\omega)a^2}$ \cite{canaliTheoryRamanScattering1992, shastryTheoryRamanScattering1990}. Firstly, we note that this interaction is quadratic in electric field, which is typically small and thus we expect this interaction to be weaker. 
	Introducing the Holstein-Primakoff operators, Fourier transforming and applying the Bogoliubov transformation, the interaction becomes
	\begin{multline}
		\hat H_{q} = \frac{S}{N_m}\sum_{\bm k\bm q\bm q'} (\bm E_{\bm q}\cdot \bm \Pi_{\bm q+\bm k})(\bm E_{\bm q'}^\dagger\cdot \bm \Pi_{\bm q+\bm k}) \hat\phi_{\bm q}^\dagger\hat\phi_{\bm q'}\times \\
		(u_{\bm k}u_{\bm q'-\bm q-\bm k} \alpha_{\bm k}\beta_{\bm q'-\bm q-\bm k} + v_{\bm k}v_{\bm q'-\bm q-\bm k} \alpha^\dagger_{-\bm k}\beta^\dagger_{-\bm q'+\bm q+\bm k}) + (\bm q \leftrightarrow\bm q'),
	\end{multline}
	where we have only kept the interactions that will generate a finite self energy at zero temperature. Here, $\bm \Pi_{\bm k}=B\sum_{ij}e^{-\ii \bm r_{ij}\cdot\bm k}\bm r_{ij}$ is the Fourier transform of the nearest-neighbor vectors $\bm r_{ij}$.

\begin{figure}
		\centering
		\includegraphics[width=0.5\columnwidth]{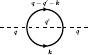}
		\caption{Feynmann diagram of the plasmon self energy resulting from the quadratic magneto-optical interaction. Dashed and solid lines indicate plasmon and magnon propagators respectively. \label{fig:sigma-raman}}
	\end{figure}

	We construct the interacting plasmon Green's function by replacing $H_c$ with $H_q$ in \cref{eq:Gint}. We can now apply Wick's theorem, which gives as the only possible plasmon self-energy the one drawn in \cref{fig:sigma-raman}. Explicitly, we have
	\begin{multline}
		\Sigma_{\bm q}(\epsilon) = \frac{N_z^2}{N_m^2}\sum_{\sigma\sigma'}\sum_{kq'} |(\bm E_{\bm q}\cdot\bm{\mathcal Y}_{\bm q,\bm k}^{\sigma\sigma'})(\bm E_{\bm q'}^*\cdot\bm{\mathcal Y}_{\bm q,\bm k}^{\sigma\sigma'})|^2 \times\\
		\mathcal G_p^{(0)}(q')\mathcal G_m^{(0)}(k)\mathcal G_m^{(0)}(q-q'-k), \label{eq:sigma-quadratic}
	\end{multline}
	where $\bm{\mathcal Y}_{\bm q,\bm k}^{\sigma\sigma'}$ is the interaction vertex and we used the shorthand notation $k\equiv(\bm k,\ii\omega_n)$. $\mathcal G_{p/m}^{(0)}$ denote the noninteracting plasmon/magnon Green's function. We do not give the full form of the interaction vertex $\bm{\mathcal Y}_{\bm q,\bm k}^{\sigma\sigma'}$ here, since we will show (see below) that the this self energy is strongly suppressed.

	Since this is a product of three noninteracting Green's functions, we do not perform the Matsubara summation here. Instead, as a first approximation, we evaluate this self energy on-shell, such that we have energy conservation at both vertices:
	\begin{equation}
		\mathcal E_{\bm q} = \mathcal E_{\bm q'} + \epsilon_{\bm q - \bm q'-\bm k} + \epsilon_{\bm k}.
	\end{equation}
	We thus have that $\mathcal E_{\bm q'}<\mathcal E_{\bm q}$. Since the plasmon dispersion is very steep compared to the magnon dispersion, $|\bm q|$ itself is very small. Therefore, the summation over $|\bm q'|$ only runs over a small part of the total phase space $q'$ [cf. the summation in \cref{eq:sigma-quadratic}], significantly suppressing the strength of this self energy. In addition, this self energy is quartic in the electric field, which will suppress its strength further. We therefore conclude that the magneto-optic interaction is too weak to give any significant correction to the plasmon self-energy.

\end{document}